\documentclass[twoside,10pt,a4paper]{newFNLstyle}
\usepackage{graphics}
\usepackage{cite}

\begin{document}

\volnumpagesyear{5}{0}{000--000}{2005}
\dates{received date}{revised date}{accepted date}

\title{GENERALIZED WIENER PROCESS
AND KOLMOGOROV'S EQUATION FOR DIFFUSION INDUCED BY NON-GAUSSIAN
NOISE SOURCE}

\authorsone{ALEXANDER DUBKOV}
\affiliationone{Radiophysics Department, Nizhny Novgorod State
University, 23 Gagarin Ave., \\ Nizhny Novgorod, 603950 Russia}
\mailingone{dubkov@rf.unn.ru}

\authorstwo{BERNARDO SPAGNOLO}
\affiliationtwo{INFM and Dipartimento di Fisica e Tecnologie
Relative, Group of Interdisciplinary Physics, Universit\`{a} di
Palermo, Viale delle Science pad. 18, I-90128 Palermo, Italy}
\mailingtwo{spagnolo@unipa.it}

\maketitle

\markboth{Generalized Wiener process and Kolmogorov's
equation...}{Dubkov and Spagnolo}

\pagestyle{myheadings}


\begin{abstract}
We show that the increments of generalized Wiener process, useful
to describe non-Gaussian white noise sources, have the properties
of infinitely divisible random processes. Using functional
approach and the new correlation formula for non-Gaussian white
noise we derive directly from Langevin equation, with such a
random source, the Kolmogorov's equation for Markovian
non-Gaussian process. From this equation we obtain the
Fokker-Planck equation for nonlinear system driven by white
Gaussian noise, the Kolmogorov-Feller equation for discontinuous
Markovian processes, and the fractional Fokker-Planck equation for
anomalous diffusion. The stationary probability distributions for
some simple cases of anomalous diffusion are derived.
\end{abstract}

\section{Introduction}

Stochastic dynamics is useful to model many biological, chemical, economical
and physical systems. The random driving forces have very different origins,
in most cases they are Gaussian white or colored noise sources, but often
these forces must be considered as non-Gaussian ones, for example, in
sensory and biological systems \cite{Bez95}. Moreover, in many physical and
biological systems a deviation of real statistics of fluctuations from
Gaussian law, leading to anomalous diffusion, is observed \cite{Cas99,Met00}%
. A suitable mathematical model to describe the anomalous
diffusion is the fractional Fokker-Planck equation, which can be
derived from different theoretical approaches
\cite{Met00,Sai97,Wes97,Jes99,Yan00}.

In this paper we show that the increments of generalized Wiener process,
representing the integral of non-Gaussian white noise, have the properties
of infinitely divisible random processes. This enables us to obtain the
characteristic functional of non-Gaussian white noise. Then, by using
functional approach to split the correlation between stochastic functionals,
we derive the Kolmogorov's equation directly from Langevin equation with
non-Gaussian white noise source. From this general equation we can obtain
the Fokker-Planck equation for continuous Markovian processes, the
Kolmogorov-Feller equation for discontinuous Markovian processes and the
fractional Fokker-Planck equation for anomalous diffusion.

\section{Generalized Wiener process}

It is well known that Wiener process $\eta \left( t\right) $ has stationary
and independent increments, and its time derivative $\xi \left( t\right)
=\dot \eta \left( t\right) $ is Gaussian white noise. The correlation
function of the increments $\eta \left( t+T\right) -\eta \left( t\right) $
depends on time difference $T$ only, and increments of non-overlapping time
intervals are statistically independent. Let us consider now a non-Gaussian
random process with stationary and independent increments which can be
called as \emph{generalized Wiener process}. We can represent the increment $%
\eta _T\equiv \eta \left( T\right) -\eta \left( 0\right) $ as
\begin{equation}
\eta_{T}=\sum\limits_{k=1}^{n}\left[ \eta\left( k\Delta t\right)
-\eta\left( \left( k-1\right) \Delta t\right) \right],
\label{incr}
\end{equation}
where $\Delta t=T/n$. Since the random variable $\eta_{T}$ can be
divided into the sum of arbitrary number of independent and
identically distributed random variables, its probability
distribution belongs to the class of infinitely divisible
distributions. Hence, we can express the logarithm of
characteristic function of the random value $\eta_{T}$ in the
L\'evy--Khinchine form \cite{Fel71}
\begin{equation}
\phi_{T}\left( u\right) \equiv\ln\theta_{T}\left( u\right) =\ln
\left\langle e^{iu\eta_{T}}\right\rangle =\int\nolimits_{-\infty}^{\infty }%
\frac{e^{iux}-1-iu\sin x}{x^{2}}\,M_{T}\left\{ dx\right\} ,
\label{Levy-Khinchine}
\end{equation}
where $M_{T}\{ dx\}$ is the canonical measure which, according to
definition, takes positive values on a finite intervals, and the
integrals
$$
M^{+}_{T}(x)=\int\nolimits_{x}^{\infty }\frac{M_{T}\left\{
dy\right\}}{y^2}\,,\qquad
M^{-}_{T}(-x)=\int\nolimits_{-\infty}^{-x}\frac{M_{T}\left\{
dy\right\}}{y^2}
$$
converge for any positive $x$ and $T$. Introducing the
non-negative density $\rho _T(x)$ of the canonical measure:
$M_{T}\{ dx\} =\rho _T(x)dx$, defined in generalized sense, we can
rewrite Eq.~(\ref{Levy-Khinchine}) as
\begin{equation}
\phi_{T}\left( u\right) =\int\nolimits_{-\infty}^{\infty }%
\frac{e^{iux}-1-iu\sin x}{x^{2}}\,\rho _T\left( x\right) dx.
\label{Levy}
\end{equation}
For two consecutive time intervals $T$ and $S$, we have
$$
\theta _{T+S}\left( u\right) =\left\langle e^{iu\eta _{T+S}}\right\rangle
=\left\langle e^{iu\left( \eta _T+\eta _S\right) }\right\rangle
=\left\langle e^{iu\eta _T}\right\rangle \left\langle e^{iu\eta
_S}\right\rangle =\theta _T\left( u\right) \theta _S\left( u\right) ,
$$
and
\begin{equation}
\label{fert}\phi_{T+S}\left( u\right) =\phi_{T}\left( u\right)
+\phi_{S}\left( u\right) .
\end{equation}
Because $u$ is an arbitrary parameter, we obtain from Eqs.~(\ref{Levy}) and (%
\ref{fert})
\begin{equation}
\rho _{T+S}\left( x\right) =\rho _T\left( x\right) +\rho _S\left(
x\right) ,\quad \forall x.
\label{Gamel}
\end{equation}
The differentiable solution of Eq.~(\ref{Gamel}), regarding parameter $T$,
is only linear one
$$
\rho _T\left( x\right) =T\rho \left( x\right) .
$$
So, from Eq.~(\ref{Levy}) we obtain
\begin{equation}
\phi_{T}\left( u\right) =T\int\nolimits_{-\infty}^{\infty}\frac
{e^{iux}-1-iu\sin x}{x^{2}}\,\rho\left( x\right) dx, \label{Ch-F}
\end{equation}
where the kernel $\rho\left( x\right) \geq0$.

\section{Functional correlation formula for non-Gaussian white noise}

The time derivative of generalized Wiener process $\xi\left(
t\right) =\dot{\eta}\left( t\right) $ is a stationary random
process and has analogy to Gaussian white noise. Now we derive the
characteristic functional of this non-Gaussian delta-correlated
noise. By definition, we have
\begin{eqnarray}
\Theta_{t}\left[  u \right]  &=& \left\langle \exp\left\{
i\int\nolimits_{0} ^{t}\xi\left(  \tau\right)  u\left( \tau\right)
d\tau\right\} \right\rangle = \left\langle \exp\left\{ i
\int\nolimits_{0}^{t}u\left(  \tau\right) d\eta\left( \tau\right)
\right\}  \right\rangle \nonumber \\
&=& \left\langle \exp\left\{
i\lim\limits_{\delta_{\tau}\rightarrow0}
\sum\limits_{k=1}^{n}u\left(  \vartheta_{k}\right)  \left[
\eta\left( \tau_{k}\right)  -\eta\left(  \tau_{k-1}\right) \right]
\right\} \right\rangle \label{Stil}  \\  &=&
\lim\limits_{\delta_{\tau}\rightarrow0}\left\langle
\prod\limits_{k=1} ^{n}\exp\left\{  iu\left(  \vartheta_{k}\right)
\left[ \eta\left( \tau _{k}\right)  -\eta\left(  \tau_{k-1}\right)
\right]  \right \} \right\rangle =
\lim\limits_{\delta_{\tau}\rightarrow0}
\prod\limits_{k=1}^{n}\theta_{\Delta\tau_{k}}\left(  u\left(
\vartheta _{k}\right)  \right), \nonumber
\end{eqnarray}
where $\vartheta _k$ is some internal point of time interval $\left( \tau
_{k-1},\tau _k\right) \,$, $\delta _\tau =\max \limits_k\Delta \tau _k$, $%
\Delta \tau _k=\tau _k-\tau _{k-1}$ ($\tau _0=0$, $\tau _n=t$),
and $\theta _{\Delta \tau _k}$ is the characteristic function of
increments. To obtain Eq.~(\ref{Stil}) we used the statistical
independence of increments of generalized Wiener process $\eta
\left( t\right) $. Further from Eqs.~(\ref{Ch-F}) and (\ref{Stil})
we obtain
\begin{eqnarray}
\Theta_{t}\left[  u\right]  &=&
\lim\limits_{\delta_{\tau}\rightarrow0}\prod
\limits_{k=1}^{n}\exp\left\{
\Delta\tau_{k}\int\nolimits_{-\infty}^{\infty }\frac{e^{iu\left(
\vartheta_{k}\right)  x}-1-iu\left(  \vartheta_{k}\right) \sin
x}{x^{2}}\,\rho\left(  x\right)  dx\right\}  \nonumber \\
&=& \exp\left\{  \lim\limits_{\delta_{\tau}\rightarrow0}\sum\limits_{k=1}%
^{n}\Delta\tau_{k}\int\nolimits_{-\infty}^{\infty}\frac{e^{iu\left(
\vartheta_{k}\right)  x}-1-iu\left(  \vartheta_{k}\right)  \sin x}{x^{2}}%
\,\rho\left(  x\right)  dx\right\} \nonumber \\
&=& \exp\left\{  \int\nolimits_{0}^{t}d\tau
\int\nolimits_{-\infty}^{\infty}\frac{e^{iu\left(  \tau\right)
x}-1-iu\left( \tau\right)  \sin x}{x^{2}}\,\rho\left(  x\right)
dx\right\} . \label{main}
\end{eqnarray}
Now we derive a useful correlation formula for stochastic functionals of
non-Gaussian white noise. We use the generalization of Furutsu-Novikov
formula \cite{Kly74}, to split the correlation between a stochastic
functional $R_{t}\left[ \xi\right] $ of non-Gaussian random process $%
\xi\left( \tau\right) $, defined on the observation interval $\tau\in\left(
0,t\right) $, and the process $\xi\left( t\right) $ itself
\begin{equation}
\left\langle \xi \left( t\right) R_t\left[ \xi +z\right]
\right\rangle =\left. \frac{\dot \Phi _t\left[ u\right] }{iu\left( t\right) }%
\right| _{\,u=\frac{\delta }{i\delta z}}\left\langle R_t\left[ \xi
+z\right] \right\rangle . \label{Kly}
\end{equation}
Here $z\left( t\right) $ is an arbitrary deterministic function,
and $\Phi _t\left[ u\right] =\ln \Theta _t\left[ u\right] $. From
Eq.~(\ref{main}) we obtain the following expression for
variational operator in Eq.~(\ref{Kly})
$$
\frac{\dot{\Phi}_{t}\left[ u\right] }{iu\left( t\right) }=\int
\nolimits_{-\infty}^{\infty}\frac{e^{iu\left( t\right)
x}-1-iu\left( t\right) \sin x}{iu\left( t\right) x^{2}}\rho\left(
x\right) dx=\int \nolimits_{-\infty}^{\infty} \frac{\rho\left(
x\right) }{x^{2}}\,dx \int\nolimits_{0}^{x}[e^{iu\left( t\right)
y}-\cos y]dy.
$$
Substituting this expression in Eq.~(\ref{Kly}) we arrive at
$$
\left\langle \xi\left( t\right) R_{t}\left[ \xi+z\right]
\right\rangle =\int\nolimits_{-\infty}^{\infty}\frac{\rho\left(
x\right) }{x^{2}}\,\,dx\int\nolimits_{0}^{x}\left[ \exp\left\{
y\frac{\delta}{\delta z\left( t\right) }\right\} -\cos y\right]
\left\langle R_{t}\left[ \xi+z\right] \right\rangle dy.
$$
By inserting the operator of functional differentiation into the average and
by putting $z = 0$, we get finally
\begin{equation}
\left\langle \xi\left( t\right) R_{t}\left[ \xi\right]
\right\rangle =\int\nolimits_{-\infty}^{\infty}\frac{\rho\left( x\right) }{%
x^{2}}\,dx\int\nolimits_{0}^{x}\left[ \left\langle \exp\left\{ y\frac{\delta}{%
\delta\xi\left( t\right) }\right\} R_{t}\left[ \xi\right]
\right\rangle -\left\langle R_{t}\left[ \xi\right] \right\rangle
\cos y\right] dy. \label{new}
\end{equation}

\section{Derivation of Kolmogorov's equation}

Let us consider now the Langevin equation with a non-Gaussian white noise $%
\xi(t)$
\begin{equation}
\dot x=f\left( x,t\right) +g\left( x,t\right) \xi \left( t\right)
. \label{Lang}
\end{equation}
By differentiating with respect to time the following expression for
probability density of the random process $x(t)$
\begin{equation}
W\left( x,t\right) =\left\langle \delta\left( x-x\left( t\right)
\right) \right\rangle ,
\label{W}
\end{equation}
and taking into account Eq.~(\ref{Lang}), we obtain
\begin{equation}
\frac{\partial W}{\partial t}= -\frac{\partial}{\partial x}%
\,\left[ f\left( x,t\right) W\right] -\frac{\partial}{\partial
x}\,\,g\left( x,t\right) \left\langle \xi\left( t\right)
\delta\left( x-x\left( t\right) \right) \right\rangle .
\label{prel}
\end{equation}
By using functional differentiation rules and following the same
procedure of Ref.~\cite{Han78}, from Eq.~(\ref{Lang}) we get
\begin{equation}
\frac{\delta}{\delta\xi\left( t\right) }\,\,\delta\left( x-x\left(
t\right) \right) =-\frac{\partial}{\partial x}\,\,g\left(
x,t\right) \delta\left( x-x\left( t\right) \right) . \label{equi}
\end{equation}
Thus, the operator $\delta/\delta\xi\left( t\right) $ of functional
differentiation with respect to the function $\delta\left( x-x\left(
t\right) \right) $ is equivalent to the ordinary differential operator $%
-\partial/\partial x\left( g\left( x,t\right) \,\right) $. Using Eq.~(\ref
{new}) in Eq.~(\ref{prel}) and taking into account Eq.~(\ref{equi}), we
obtain the Kolmogorov's equation for nonlinear system (\ref{Lang}) driven by
non-Gaussian white noise
\begin{equation}
\frac{\partial W}{\partial t}=-\frac{\partial \left[
f(x,t)W\right] }{\partial x}+\int\nolimits_{-\infty }^\infty
\frac{\rho \left( z\right) }{z^2}\left[ \exp \left\{ -z\frac
{\partial }{\partial x}g\left( x,t\right) \right\} -1+\sin z\frac
{\partial }{\partial x}g\left( x,t\right) \right] dzW,
\label{gene}
\end{equation}
which is the main result of our paper. By series expansion of the
exponential operator in Eq.~(\ref{gene}), we have
\begin{equation}
\frac{\partial W}{\partial t}=-\frac{\partial}{\partial x}\,\left[
f\left( x,t\right) W\right]
+\sum\limits_{n=1}^{\infty}\frac{\left( -1\right)
^{n}A_{n}}{n!}\left[ \frac{\partial}{\partial x}\,\,g\left(
x,t\right) \right] ^{n}W, \label{Han}
\end{equation}
where
\begin{equation}
A_1=\int\nolimits_{-\infty }^\infty \frac{z-\sin z}{z^2}\,\rho
(z)\,dz,\quad \quad A_n=\int\nolimits_{-\infty }^\infty
z^{n-2}\rho \left( z\right) dz,\quad n\geq 2. \label{A-n}
\end{equation}
The equation (\ref{Han}) is valid if all integrals in Eq.~(\ref{A-n})
converge. It coincides with the result obtained in Ref.~\cite{Han78} for
Poissonian white noise $\xi (t)$, if we interpret the coefficients $A_n$ as
the moments of random jumps of $x(t)$. Moreover, this equation can be recast
in the Kramers-Moyal form
\begin{equation}
\frac{\partial W}{\partial t}=\sum\limits_{n=1}^{\infty}\frac{%
\left( -1\right) ^{n}}{n!}\frac{\partial^{n}}{\partial
x^{n}}\,\left[ K_{n}\left( x,t\right) W\right] , \label{K-M}
\end{equation}
where $K_{n}\left( x,t\right) $, $n=1,2,\ldots$ are the kinetic coefficients.

\section{Particular cases of non-Gaussian noise sources}

Following Ref.~\cite{Fel71} we analyze different kernel functions
$\rho (x)$ to obtain particular cases of Kolmogorov's equation
(\ref{gene}), related to different non-Gaussian noise sources.

(a) As a first simple case we consider a Gaussian white noise $\xi (t)$. The
corresponding kernel function is $\rho \left( x\right) =2D\delta \left(
x\right) $. Therefore we obtain the ordinary Fokker-Planck equation
\begin{equation}
\frac{\partial W}{\partial t}=-\frac \partial {\partial x}\,\left[
f\left( x,t\right) W\right] +D\frac \partial {\partial
x}\,\,g\left( x,t\right) \frac \partial {\partial x}\,\,g\left(
x,t\right) W. \label{FP}
\end{equation}

(b) For Poisson kernel $\rho \left( x\right) =B\,\delta \left(
x-q\right) $, Eq.~(\ref{gene}) becomes
\begin{equation}
\frac{\partial W}{\partial t}=-\frac \partial {\partial x}\left[
f\left( x,t\right) +\frac {Bg\left( x,t\right) \left( q-\sin
q\right)}{q^2}\right] W +B\sum\limits_{n=2}^\infty \frac{\left(
-1\right) ^nq^{n-2}}{n!}\left[ \frac \partial {\partial x}g\left(
x,t\right) \right] ^nW. \label{Pois}
\end{equation}

(c) When the increment $\eta _T$ of generalized Wiener process has
the gamma distribution with exponent $\mu $, the kernel is $\rho
\left( x\right) =\mu xe^{-ax}$ $\left( x\geq 0\right) $. As a
result, after integration we find from Eq.~(\ref{gene})
\begin{equation}
\frac{\partial W}{\partial t}=-\frac {\partial}{\partial
x}\,\left[ f\left( x,t\right) -\mu g\left( x,t\right)
\mathrm{arc\,cot} \,\,a\right] W -\mu \ln \left[ 1+\frac 1a\,\frac
\partial {\partial x}\,\,g\left( x,t\right) \right] W.\nonumber
\end{equation}

(d) For one-side $\alpha $-stable L\'evy distributions the kernel
function is $\rho \left( x\right) =\rho _0\,x^{1-\alpha }$ $\left(
x\geq 0\right) $, and we get
\[
\frac{\partial W}{\partial t}=-\frac{\partial}{\partial x}\,\left[
f\left( x,t\right) W\right]
\]
\[
+\rho_{0}\int\nolimits_{0}^{\infty}\frac{dz}{z^{\alpha +1}}\left[
\exp\left\{
-z\frac{\partial}{\partial x}\,\,g\left( x,t\right) \right\} -1+\frac{\partial}{%
\partial x}\,\,g\left( x,t\right) \sin z\right] W\left( x,t\right) .
\]
After rearrangements we obtain
\begin{equation}
\frac{\partial W}{\partial t}=-\frac{\partial \left[
f(x,t)W\right]}{\partial x}
+\frac{\rho_{0}}{\alpha}\,\frac{\partial }{\partial x}\,\,g \left(
x,t\right)
\int\nolimits_{0}^{\infty}\frac{dy}{y^{\alpha}}\left[ \cos y-\exp\left\{ -y%
\frac{\partial }{\partial x}\,\,g\left( x,t\right) \right\}
\right]W. \label{frac}
\end{equation}
This equation is the fractional Fokker-Planck equation, and
describes anomalous diffusion in the form of asymmetric L\'evy
flights.

(e) For additive driving noise $\xi\left( t\right) $, $g\left( x,t\right) =1$
in Eq.~(\ref{Lang}), and the exponential operator in Eq.~(\ref{gene})
reduces to the space shift operator. As a result, we find
\begin{equation}
\frac{\partial W}{\partial t}=-\frac \partial {\partial x}\left[
f\left( x,t\right) W\right] +\int\nolimits_{-\infty }^\infty
\frac{\rho
\left( z\right) }{z^2}\left[ W\left( x-z,t\right) -W\left( x,t\right) +\sin z%
\frac{\partial W\left( x,t\right) }{\partial x}\right] dz.
\label{add}
\end{equation}
Equation (\ref{add}) is similar to the Kolmogorov-Feller equation
for purely discontinuous Markovian processes \cite{Sai97,Kam04}
\begin{equation}
\frac{\partial W}{\partial t}=\nu \int\nolimits_{-\infty }^\infty
Q\left( x-z\right) W\left( z,t\right) dz-\nu W\left( x,t\right) ,
\label{Kol}
\end{equation}
where $Q\left( x\right) $ is probability density of jumps step, and $\nu $
is the constant mean rate of jumps. Putting $f\left( x,t\right) =0$ and
comparing Eq.~(\ref{add}) with Eq.~(\ref{Kol}) we obtain the kernel function
for this case
\begin{equation}
\rho \left( x\right) =\nu x^2Q\left( x\right) .
\label{conn}
\end{equation}
For non-Gaussian driving force $\xi \left( t\right) $, with
symmetric $\alpha $-stable L\'evy distribution, the kernel
function is $\rho \left( x\right) =D\left| x\right| ^{1-\alpha }$.
As a result, Eq.~(\ref{add}) takes the following form
\begin{equation}
\frac{\partial W}{\partial t}=-\frac{\partial}{\partial x}%
\,\left[ f\left( x,t\right) W\right] +D\int\nolimits_{-\infty}^{\infty}\frac{%
W\left( z,t\right) -W\left( x,t\right) }{\left\vert x-z\right\vert
^{\alpha+1}}\,\,dz \label{anom diff}
\end{equation}
and describes the anomalous diffusion in form of symmetric L\'evy
flights.

(f) Finally, for multiplicative noise $\xi\left( t\right) $ with
$g\left( x,t\right) =x$, the operator in Eq.~(\ref{gene}) reduces
to a scaling operator, and the generalized Kolmogorov's equation
becomes
\begin{equation}
\frac{\partial W}{\partial t}=-\frac \partial {\partial x}\,[
f\left( x,t\right) -Kx]\,W +\int\nolimits_{-\infty }^\infty
\frac{\rho \left( z\right) }{z^2}\left[ W\left( e^{-z}x,t\right)
-W\left( x,t\right) \right] dz, \label{param}
\end{equation}
where
$$
K=\int\nolimits_{-\infty }^\infty \frac{\rho \left( z\right) \sin
z}{z^2}\,\,dz.
$$

\section{Simple examples of superdiffusion}

Now we derive the stationary probability distribution $W_\infty (x)$ for
some simple cases of anomalous diffusion in a fixed potential $U(x)$ and in
overdamped regime (see also \cite{Jes99,Che04,Kal04}). The Langevin equation
for the particle displacement $x(t)$ reads
\begin{equation}
\dot x=-U^{\prime }\left( x\right) +\xi \left( t\right) ,
\label{Langevin}
\end{equation}
where $\xi \left( t\right) $ is a symmetric $\alpha $-stable
L\'evy process. The corresponding Kolmogorov's equation is
Eq.~(\ref{anom diff}), with $f(x,t)=-U^{\prime }(x)$. The
stationary probability distribution $W_\infty (x)$ can be found
from the following equation
\begin{equation}
\frac d{dx}\,\left[ U^{\prime }\left( x\right) W_\infty \right]
+D\int\nolimits_{-\infty }^\infty \frac{W_\infty \left( z\right)
-W_\infty \left( x\right) }{\left| x-z\right| ^{\alpha
+1}}\,\,dz=0. \label{SPD}
\end{equation}

a) For uniform potential well with two reflecting boundaries at $x=\pm L$ we
get, from Eq.~(\ref{SPD}), the uniform stationary probability density
$$
W_\infty \left( x\right) =\frac 1{2L}
$$
as for ordinary Brownian diffusion.

b) For linear system with parabolic potential $U\left( x\right) =\beta
x^{2}/2$, by Fourier transform of Eq.~(\ref{SPD}), we obtain the asymptotic
characteristic function of the process $x(t)$ as \cite{Jes99}
\begin{equation}
\theta _\infty \left( k\right) =\exp \left\{ -\frac{2\Gamma
\left( -\alpha \right) \cos \left( \pi \alpha /2\right) }{\alpha \beta }%
\left| k\right| ^\alpha \right\} . \label{parabolic}
\end{equation}

c) For symmetric potential $U(x)=\gamma |x|$, after Fourier transform of
Eq.~(\ref{SPD}) we find
\begin{equation}
\gamma kG\left( k\right) +2D\left\vert k\right\vert
^{\alpha}\Gamma\left( -\alpha\right) F\left( k\right) \cos
\frac{\pi\alpha}{2} =0, \label{V-shape}
\end{equation}
where
$$
F\left( k\right) =\int\nolimits_{0}^{\infty}W_{\infty}\left(
x\right) \cos kx\,dx,\qquad G\left( k\right)
=\int\nolimits_{0}^{\infty}W_{\infty}\left( x\right) \sin kx\,dx.
$$
Functions $F\left( k\right) $ and $G\left( k\right) $ are connected by
Hilbert formula
\begin{equation}
G\left( k\right) =\frac 1\pi \,v.p.\int\nolimits_0^\infty \frac{%
F\left( q\right) }{k-q}\,\,dq, \label{Hilbert}
\end{equation}
where $v.p.$ means the principal value of integral. After
substitution of Eq.~(\ref{Hilbert}) in Eq.~(\ref{V-shape}) we
obtain finally the following integral equation for cosine Fourier
transform of stationary probability density
\begin{equation}
v.p.\int\nolimits_0^\infty \frac{F\left( q\right) }{k-q}\,\,dq+\frac{%
2\pi D}{\gamma k}\left| k\right| ^\alpha \Gamma \left( -\alpha
\right) F\left( k\right) \cos \frac{\pi \alpha }{2}=0\, .
\label{last}
\end{equation}

d) For quartic potential $U\left( x\right) =\beta x^4/4$, from
Eq.~(\ref{SPD}) we obtain the following differential equation for
stationary characteristic function
\begin{equation}
\beta k\,\frac{d^3\theta _\infty (k)}{dk^3}-2D\left| k\right|
^\alpha \Gamma \left( -\alpha \right) \theta _\infty (k)\cos
\frac{\pi \alpha }{2}=0. \label{quartic}
\end{equation}
Equations (\ref{last}) and (\ref{quartic}) can be solved numerically.

\section{Conclusions}

Starting from the \emph{generalized Wiener process}, whose increments have
the probability distribution belonging to the class of infinitely divisible
distributions, we obtained the new correlation formula for stochastic
functionals. This allows us to derive the Kolmogorov's equation (\ref{gene})
for diffusion, induced by non-Gaussian white noise source, directly from
Langevin equation. The Fokker-Planck equation for ordinary Brownian motion,
the Kolmogorov-Feller equation for purely discontinuous Markovian processes
and the fractional Fokker-Planck equation for anomalous diffusion are
obtained as particular cases of the Kolmogorov's equation (\ref{gene}). This
equation can be used for determining the stationary probability
distributions, probability distributions of first-passage times and other
statistical characteristics of nonlinear dynamical systems.

\section*{Acknowledgements}

This work has been supported by INTAS Grant 2001-0450, MIUR, INFM, by
Russian Foundation for Basic Research (project 02-02-17517), by Federal
Program "Scientific Schools of Russia" (project 1729.2003.2), and by
Scientific Program "Universities of Russia" (project 01.01.020).

\end{document}